\documentclass[%
 reprint,
superscriptaddress,
,amssymb,
 aps,
]{revtex4-2}

\usepackage{amsmath}
\usepackage{graphicx}
\usepackage{dcolumn}
\usepackage{bm}
\usepackage{soul}
\usepackage[colorlinks=true,allcolors=blue]{hyperref}
\usepackage{xcolor}
\newcommand{\p}[1]{\textbf{#1}}
\newcommand\rev[1]{\textcolor{black}{#1}}

\begin{document}

\title{Dumbbell dimer dynamics in three-dimensional chiral fluids}

\author{Michalis Chatzittofi}%
\email{mike.chatzittofi@ds.mpg.de}
\affiliation{%
 Max Planck Institute for Dynamics and Self-Organization, G{\"o}ttingen 37073, Germany
}%
\author{Yuto Hosaka}
\email{yuto.hosaka@ds.mpg.de}
\affiliation{%
 Max Planck Institute for Dynamics and Self-Organization, G{\"o}ttingen 37073, Germany
}%

\date{\today}

\begin{abstract}
We study the emergent orientational dynamics of a dumbbell dimer -- two asymmetric monomers connected by a linking spring -- in a three-dimensional chiral environment with odd viscosity. 
In classical systems with conserved parity symmetry, reciprocal oscillations of a dimer do not lead to rotational motion. 
Here, through an analytical calculation,  we find that the presence of chirality in the system induces rotational dynamics as a function of the expansion/contraction of the dimer.
By incorporating thermal fluctuations, we further find that the rotational diffusivity is affected by the coupling between conformational fluctuations and rotational motion.
Our results provide insights into problems where the parity symmetry is broken and can be used as a building block to study similar models at the collective level.
These problems include multi-component molecular machines in odd-viscous fluids and systems with charged polymers where oddity is present through external magnetic fields.
\end{abstract}

\maketitle
\section*{Introduction}

Recent developments in the field of active matter have revealed that the activity of various microscopic agents violates the classical symmetries in physical systems.
These violations include detailed balance and time-reversal and rotational symmetries, and they lead to novel phenomena at large scales, such as emergent collective behavior and swarming of active constituents~\cite{PhysRevX.12.010501, Gompper2025}. 
Microscopically, an individual agent, be it living or artificial, exhibits autonomous motion by harnessing chemical energy from its environment. Examples span at multiple scales, ranging from molecular proteins in active biological systems~\cite{battle2016broken} to microorganisms swimming in aqueous environments and driven colloidal systems~\cite{ignes2022experiments}.

Despite their differences in scale and variation, local force generation plays an important role in understanding the underlying collective phenomena and can be taken into account by coarse-graining the flow induced by each active agent~\cite{aditi2002hydrodynamic, mikhailov2015, thampi2016active}. For example, the diffusion enhancement observed in cellular environments has been attributed to the activity of individual molecular proteins or enzymes~\cite{guo2014probing, parry2014}. Among the various proposed mechanisms for modeling such a local force generation, the dumbbell dimer or force dipole model is widely used as a minimum description of an active agent~\cite{mikhailov2015, Illien2017}. 
The model typically consists of two subunits connected by a rigid shaft or an elastic spring that allows for shape fluctuations and diffusive motion.
Such a simple description is achieved through the dimensional reduction of an active protein with a large number of degrees of freedom.
For enzymatic molecules, size changes mimic the conformational motions in chemical reactions and are known to serve as a stirring function for cellular metabolism~\cite{losa2022perspective}. When immersed in aqueous environments, the cyclic or shaking motion induced by dimers acts as a hydrodynamic force dipole in the far field, which is the leading contribution to the flow field induced by microswimmers~\cite{lauga2016bacterial}. Motivated by diffusion enhancement reported in experiments, many theoretical works have been performed~\cite{golestanian2015}, which includes the modeling of an enzymatic dimer~\cite{Illien2017, adeleke2019, tyagi2021effects} and the statistical properties of collective dimers in dilute~\cite{mikhailov2015, suma2014dynamics} and in crowded environments~\cite{dennison2017diffusion, koyano2020diffusion, klett2021non}.

In addition to modeling an individual active particle, the dimer model has also been used to study the rheological properties of soft media via microrheology techniques.
Typically, a single colloidal particle is used as a probing tracer to explore various systems~\cite{crocker2000two, furst2017microrheology}, such as bacterial baths~\cite{wu2000particle} and living cells~\cite{ebata2023activity, muenker2024accessing}.
However, tracer objects with anisotropy or multi-components can reveal more complicated and mesoscopic structures of the media. For instance, spherical dimers have been employed to study the effect of viscoelasticity on propulsion~\cite{sahoo2019enhanced}, the dynamics in a Newtonian fluid with viscosity gradient~\cite{liebchen2018viscotaxis}, and the recoil dynamics in a relaxing environment~\cite{krishna2023memory}.

Recently, the transport and dynamics of microparticles in chiral environments, which themselves break parity symmetries, has gained much attention and the macroscopic consequence of the self-spinning fluid components or particles in chiral motion has been studied~\cite{torrik2021dimeric, liebchen2022chiral, mecke2024emergent}.
Within the continuum level description, such fluid systems with broken parity exhibit intriguing chiral or ``odd" responses, which are of current interest in the field~\cite{banerjee2017, hosaka2022nonequilibrium, Fruchart2023}. 
From the symmetry argument, chiral systems naturally acquires an additional dissipationless transport coefficient called \textit{odd viscosity}~\cite{khain2022}. The emergence of the new quantitative measure has motivated and stimulated work proposing a new protocol to quantify active rheological properties experimentally~\cite{zhao2021}. The continuum description characterized by the odd viscosity has revealed various striking effects, such as transverse responses in the viscous, low-Reynolds-number regime~\cite{lapa2014, ganeshan2017, khain2024trading, hosaka2023lorentz, aggarwal2023thermocapillary, lier2024slip, matus2024molecular}, as well as peculiar flocking and pattern formations observed at intermediate and high Reynolds numbers~\cite{deWit2024, chen2024self}. 
Nevertheless, regardless of recent advances in theoretical studies, the hydrodynamic interactions and resulting multi-particle dynamics in chiral fluids have yet to be fully explored~\cite{fruchart2023odd}. 
This is largely because exact analytical solutions of the fundamental hydrodynamic problem involving the response due to a point force and finite-sized particles remained scarce until just recently, especially in three-dimensional chiral fluids~\cite{everts2024dissipative, hosaka2024chirotactic}, compared to two-dimensional counterparts~\cite{ganeshan2017, hosaka2021hydrodynamic, daddi2025analytical, daddi2025hydrodynamic}.
In two-dimension, the chiral systems are compatible with fluid isotropy~\cite{avron1998} and can be relatively easily treated within the analytical framework using the complex notation~\cite{avron1998, lapa2014}, whereas the emergent anisotropy leads to the nontrivial coupling between the spatial degrees of freedom in three-dimensional chiral systems.

\begin{figure}[t]
\centering
\includegraphics[width=0.8\linewidth]{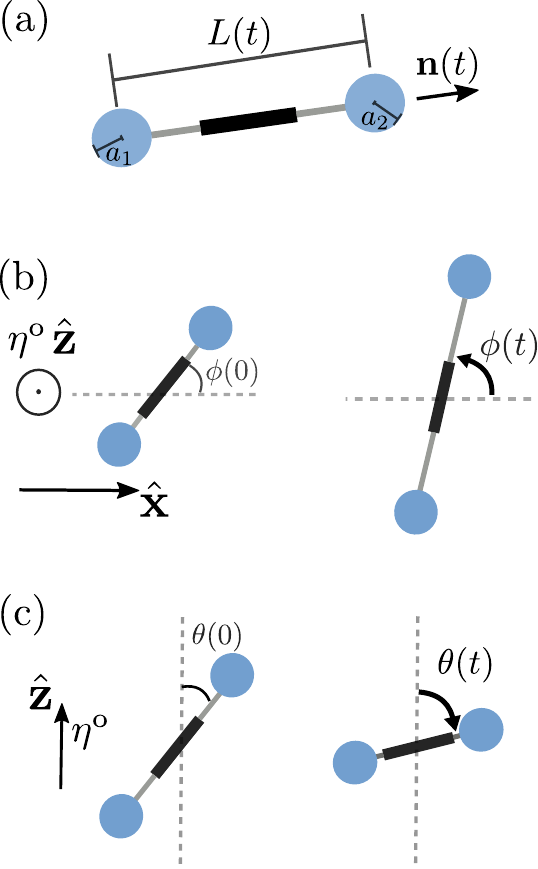}
\caption{\rev{
Schematic of a dumbbell dimer immersed in a three-dimensional chiral fluid with an even (shear) viscosity and an odd viscosity, $\eta^{\rm e}$ and $\eta^\mathrm{o}$. The odd viscosity is assumed to be positive. The axis of the odd viscosity is assumed to be in the $\hat{\p{z}}$-direction, while the even viscosity is isotropic and thus does not have any preferred direction in the fluid.
(a) The geometry of a dimer whose spherical domains have radii $a_1$ and $a_2$.
The dimer length as well as the dimer orientation vary in time, which are characterized by the scalar function $L(t)$ and the unit vector $\p{n}(t)$, respectively. 
The dimer length undergoes expansions or contractions due to thermal noise or prescribed motion of the connecting shaft, which leads to the three dimensional rotational motion of the dimer.
(b) To linear order in the odd viscosity, the expansion of the dimer arm creates an anti-clockwise rotation about the $\hat{\p{z}}$-axis by increasing the azimuthal angle $\phi(t)$ measured from the $x$-axis [see also Eq.~\eqref{eq:phisol} and Fig.~\ref{fig:fig2}(b)].
(c) To quadratic order in the odd viscosity, in addition to the transient precession observed in panel (b), the contraction of the dimer causes the reorientation towards the $(x,y)$ plane by increasing the polar angle $\theta(t)$ measured from the $z$-axis [see also Eq.~\eqref{eq:thetasol} and Fig.~\ref{fig:fig2}(a)].}
}
\label{fig:fig1}
\end{figure}

In this work, we study the dynamics of a dumbbell dimer immersed in unbounded three-dimensional chiral fluids \rev{(Fig.~\ref{fig:fig1})}.
The dimer is either active by changing its conformation cyclically or passive, undergoing thermal fluctuations.
Employing the continuum description of chiral systems characterized by odd viscosity, we derive the spatial equations of the dimer motion and show that the mobility of velocity-force relations obtains a non-symmetric structure in the exchange of spatial index.
This suggests that odd viscosity naturally gives rise to an odd mobility in the spatial dynamics of particles and creates an intriguing rotational behavior~\rev{[Fig.~\ref{fig:fig1}(b,c)]}, prohibited in conventional fluids.
Using the mobility relations in fluids with odd viscosity, we study the two cases: the interdomain hydrodynamic interaction up to the leading order in odd viscosity and the higher order effect.
Moreover, by extending the odd-viscous system in the presence of the thermal noise, we investigate the interplay between the fluctuation and oddity and how it affects the rotational dynamics of the dimer.

\section*{Model of a passive and active dimer in viscous fluids}

We consider an axisymmetric dimer submerged in a three-dimensional incompressible chiral fluid with parity violation \rev{(Fig.~\ref{fig:fig1})}.
This section recapitulates the concept of odd viscosity, a general measure that characterizes chiral fluid environments~\cite{banerjee2017, fruchart2023odd}.
At low Reynolds number, where inertial is negligible, the fluid property is governed by the force balance equation
\begin{align}
    \partial_j\sigma_{ij}=0
    \label{eq:balance}
    ,
\end{align}
together with the incompressibility condition $\partial_iv_i=0$.
Here $\boldsymbol{\sigma}$ represents the fluid stress tensor, $\partial_i$ is the three-dimensional differential operator with $i,j,k = \{x,y,z\}$, and the summation over repeated indices is assumed.
The important relation here is the constitutive relation, which relates the stress field to the gradient of the velocity field:
\begin{align}
    \sigma_{ij} = \eta_{ijk\ell}\partial_\ell v_k,
\end{align}
where $\eta_{ijk\ell}$ is the fourth-rank viscosity tensor and includes any fluid property.
In order to introduce a minimum continuum description of chiral fluids, we assume the classical assumptions, such as the momentum conservation and the absence of the response to the vorticity, which are given by the equalities $\sigma_{ij}=\sigma_{ji}$ and $\eta_{ijk\ell}=\eta_{ij\ell k}$, respectively.
To take into account the chirality in fluids, one should impose the left- and right-handedness on the viscosity tensor by introducing the parity violation around some axis, while rotational invariance about the direction can still be assumed~\cite{khain2022, fruchart2023odd}.
Throughout this paper, we choose the axis of odd viscosity along the $z$-direction in the lab frame, without loss of generality.
In this case, the viscosity tensor accounting for parity violation is simply expressed with two transport coefficients, shear and odd viscosities, $\eta^{\rm e}$ and $\eta^{\rm o}$~\cite{khain2022, markovich2021}
\begin{align}\label{eq:OV}
    \eta_{ijk\ell}
    &=
    \eta^{\rm e}
    \left(
    \delta_{ik}\delta_{j\ell}
    +
    \delta_{i\ell}\delta_{jk}
    -
    \frac{2}{3}
    \delta_{ij}\delta_{k\ell}
    \right)\nonumber\\
    &+
    \frac{\eta^{\rm o}}{2}
    \left(
    \epsilon_{j\ell z}\delta_{ik}
    +
    \epsilon_{i\ell z}\delta_{jk}
    +
    \epsilon_{ik z}\delta_{j\ell}
    +
    \epsilon_{jkz}\delta_{i\ell}
    \right)
    ,
\end{align} 
where $\delta_{ij}$ is the Kronecker delta and $\epsilon_{ijk}$ is the three-dimensional Levi-Civita tensor with $\epsilon_{xxz}=\epsilon_{yyz}=0$ and $\epsilon_{xyz}=-\epsilon_{yxz}=1$.
Notice that the tensorial structure with $\eta^\mathrm{o}$ is antisymmetric or odd under $ij\leftrightarrow k\ell$, reflecting the chiral nature of fluids, unlike symmetric components for the classical counterpart $\eta^\mathrm{e}$.
This anti-symmetric nature suggests that $\eta^{\rm o}$ can also be negative, as the associated tensor does not contribute to the fluid energy dissipation~\cite{khain2022}.

In an unbounded fluid described by the momentum conservation~\eqref{eq:balance} and the viscosity tensor~\eqref{eq:OV}, we consider a dumbbell dimer composed of two spherical subunits with radius $a_1$ and $a_2$, connected via a shaft that can vary its length.
The position of each sphere is denoted by the three-dimensional vector $\p{r}_\alpha$ ($\alpha=1,2$).
Denoting the velocity of each sphere by $\dot{\p{r}}_\alpha=d \p{r}_\alpha/dt$ and the force acting on each subunit by $\p{f}_\alpha$, the overdamped equations of motion can be written as
\begin{align}
    \dot{\p{r}}_\alpha &= \sum_{\beta=1}^2\p{M}_{\alpha\beta} \cdot \p{F}_\beta
    .
\end{align}
Here the mobility matrix $\p{M}_{\alpha\beta}$ accounts for the hydrodynamic contributions on the subunits exerted by the surrounding fluid.
The diagonal mobility tensor $\p{M}_{\alpha\alpha}$ is associated with the hydrodynamic force acting on a each sphere,
while the off-diagonal mobility  tensor $\p{M}_{\alpha\beta}$ is related to the hydrodynamic interactions between the subunits.
The off-diagonal components or Green's function of the linear hydrodynamic equations further satisfy the relation $\partial_j M_{\alpha\beta,ij}=0$ from the incompressibility condition of the velocity field.
In a conventional fluid with shear viscosity alone, theses two contributions are, respectively, attributed to Stokes' law for a rigid sphere and the Oseen tensor, which is the solution of the forced Stokes equation~\cite{happel2012low}.
In a fluid with odd viscosity, however, the chirality in the fluid gives rise to the anti-symmetric components in the mobility matrix, leading to asymmetric relations under the exchange of the spatial index, i.e., $(M_{\alpha\beta})_{ij}\neq (M_{\alpha\beta})_{ji}$~\cite{hosaka2021hydrodynamic, jia2022incompressible, yuan2023stokesian, hosaka2023lorentz, lier2024slip, everts2024dissipative, khain2024trading}.
This condition holds for both the diagonal and off-diagonal mobility elements.
Note that symmetric relations still hold under the particle notation, $\p{M}_{\alpha\beta} = \p{M}_{\beta\alpha}$.

\rev{
When the distances between the spheres are sufficiently larger than their sizes $(a_\alpha\ll|\mathbf{r}_\alpha-\mathbf{r}_\beta|\equiv r_{\alpha\beta})$, $\mathbf{M}_{\alpha\beta}$ can be approximated as~\cite{everts2024dissipative}
\begin{align}
    \mathbf{M}_{\alpha\alpha}
    =&
    \frac{1}{24\pi\eta^{\rm e}a_\alpha}
    \left[
    \left(\frac{m_-\lambda^2}{4} +4\right)(\mathbf{I}-\hat{\mathbf{z}}\hat{\mathbf{z}})
    \right.\nonumber\\
    &\left.+\left(\frac{m_+\lambda^2}{2}+4\right)\hat{\mathbf{z}}\hat{\mathbf{z}}
    +f\lambda \boldsymbol{\epsilon}\cdot\hat{\mathbf{z}}
    \right],
    \label{eq:Maa}
    \\
    \mathbf{M}_{\alpha\beta}
    =&
    \frac{1}{4\pi\eta^{\rm e}(r_{\alpha\beta}+r_\lambda)}\nonumber\\
    &\times\Bigg[
    \mathbf{I}+\mathbf{nn}
    -\frac{\lambda r_{\alpha\beta}}{2r_\lambda}
    \boldsymbol{\epsilon}\cdot\left(\hat{\mathbf{z}}-(\mathbf{n}\cdot\hat{\mathbf{z}})\mathbf{n}\right)\nonumber\\
    &-\left(1-\frac{r_{\alpha\beta}}{r_\lambda}\right)
    \left(\mathbf{nn}+\frac{(\mathbf{n}\times\hat{\mathbf{z}})(\mathbf{n}\times\hat{\mathbf{z}})}{1-(\p{n}\cdot \hat{\p{z}})^2}\right)
    \Bigg],
    \label{eq:Mab}
\end{align}
where $r_\lambda=r_{\alpha\beta}\sqrt{4+\lambda^2[1-(\p{n}\cdot \hat{\p{z}})^2]}/2$ and $\displaystyle\p{n} = (\p{r}_\alpha-\p{r}_\beta)/r_{\alpha \beta}$ denotes a unit vector along the axis of a dumbbell dimer [Fig.~\ref{fig:fig1}(a)].
In the above, the scaling functions of $\lambda$ are defined as $m_+=f+g$ and $m_-=f-g$ with $\displaystyle f=-12\left(2\psi+\lambda\right)\lambda^{-3}$, $\displaystyle g=4(1+f)\lambda^{-2}$, and $\displaystyle\psi=\arcsin\bigl(-\lambda(4+\lambda^2)^{-1/2}\bigr)$~\cite{everts2024dissipative}.
Expression~\eqref{eq:Maa} corresponds to the odd Stokes mobility, while Eq.~\eqref{eq:Mab} describes the hydrodynamic interaction governed by Green's function, which is the solution of a point force acting on an unbounded fluid with $\eta^{\rm e}$ and $\eta^{\rm o}$.
These expressions are valid for any odd-to-even viscosity ratio, $\lambda=\eta^{\rm o}/\eta^{\rm e}$.
Note that the hydrodynamic effects on the subunits due to the presence of the connecting shaft are assumed to be negligible~\cite{najafi2004}, so that the interaction can be described by the Green's function of a point force, as in Eq.~\eqref{eq:Mab}.}

This setup is generic and includes both the cases for passive and active dimers depending on the nature of the forces. 
Passive forces are typically the gradient of a conservative two-pair potential minimally modeled by a harmonic potential. 
On the other hand, active forces could arise from non-equilibrium mechanochemical mechanisms that transduce chemical energy (typically through ATP hydrolysis) into mechanical active conformational oscillations~\cite{PhysRevLett.127.208103, Chatzittofi2025} that was recently reported in experiments of active nanomotors~\cite{Tang2025Jan}.

\section*{Orientational dynamics of active dimer}

To determined the dynamics of an active dumbbell dimer, it is convenient to introduce the relative frame of reference where $\p{L} = \p{r}_2 - \p{r}_1$ and $\p{R} = (\p{r}_1+\p{r}_2)/2$. Additionally, we write the relative vector as $\p{L}= L \p{n}$ where $L$ is the length of the dimer and $\p{n}$ its orientation.
In spherical coordinates, they are expressed as $\p{n} = \sin\theta \cos\phi \hat{\p{x}}+ \sin\theta \sin\phi  \hat{\p{y}} + \cos\theta  \hat{\p{z}}$ where $\theta$ is the polar angle measured from the $z$-axis (in the laboratory frame) and $\phi$ is the azimuthal angle around $\hat{\mathbf{z}}$ \rev{(see Fig.~\ref{fig:fig1})}.
\rev{Unless otherwise stated, we consider situations in which dimers deform their shapes in a time-reversal manner, i.e., undergoing reciprocal motion.
In general, however, the dimer model is not limited to this type of deformation and can also exhibit any deformation specified by the dimer length, $L(t)$.}
Since the considered active dimer autonomously changes its conformation without external forces, one should impose the force-free and torque-free conditions on it.
This leads to the relations $\p{F}_2=-\p{F}_1\equiv\p F$ and $\p{n} \times \p{F} =\p 0$.

\rev{
By using the relations $\dot{\mathbf{L}}=(\mathbf{M}_{11}+\mathbf{M}_{22}-2\mathbf{M}_{12})\cdot\mathbf{F}$ and $\dot{\mathbf{R}}=(\mathbf{M}_{22}-\mathbf{M}_{11})\cdot\mathbf{F}/2$ and applying the torque-free condition to them, the dynamics of the dimer conformation simplifies to
\begin{align}
    \dot{\mathbf{L}}
    &=
    \mathbf{M}_+\cdot\mathbf{F}\nonumber
    \label{eq:Ldotgen}\\
    &-
    \frac{1}{2\pi\eta^{\rm e}(r_{12}+r_\lambda)}
    \left[
    \left(1+\frac{r_{12}}{r_\lambda}\right)
    \mathbf{I}
    -\frac{\lambda r_{12}}{2r_\lambda}
    \boldsymbol{\epsilon}\cdot\hat{\mathbf{z}}
    \right]\cdot\mathbf{F},\\
    \dot{\mathbf{R}}
    &=
    -\frac{1}{2}\mathbf{M}_-\cdot\mathbf{F}
    \label{eq:Rdotgen},
\end{align}
where we have defined the self-mobility difference as 
\begin{align}
    &\mathbf{M}_\pm
    =
    \frac{1}{24\pi\eta^{\rm e}}
    \left(
    \frac{1}{a_1}\pm\frac{1}{a_2}
    \right)\nonumber\\
    &\times
    \left[
    \biggl(
    \frac{m_-\lambda^2}{4} 
    +
    4
    \biggr)\mathbf{I}
    +
    \frac{\lambda^2}{2}
    \biggl(
    m_+
    -
    \frac{m_-}{2}
    \biggr)
    \hat{\mathbf{z}}\hat{\mathbf{z}}
    +
    f\lambda \boldsymbol{\epsilon}\cdot\hat{\mathbf{z}}
    \right].
\end{align}
In the presence of odd viscosity $(\eta^{\rm o}\neq0)$, the self-mobility causes velocity not only along the dimer director $\mathbf{n}$, but also in the $z$- and azimuthal directions.
Notice that the transverse force along $\hat{\mathbf{z}}$ is absent in the hydrodynamic interaction described by the Green's function.
}

To investigate the effect of the fluid chirality on the dynamics of an active dimer, we first study the case where the magnitude of the odd viscosity is much smaller than shear viscosity $(\eta^{\rm o}/\eta^{\rm e}\ll1)$.
Even in this leading-order limit, we will see the emergence of the anti-symmetric mobility tensor and it gives rise to the novel dynamics of the dumbbell dimer, which is completely different from that in classical fluids.
We then analyze the higher-order odd viscosity effect on the behavior of dimers.

\subsection*{Linear order effect of odd viscosity}


\rev{
Starting from the velocity-force equations~\eqref{eq:Ldotgen} and \eqref{eq:Rdotgen} and expanding them to the first order in $\lambda=\eta^{\rm o}/\eta^{\rm e}$, we obtain the equations of motion, which are given by}
\begin{align}
     \dot{\p{L}} =& \Bigg[\frac{1}{6\pi\eta^\mathrm{e}}\left(\frac{1}{a_1}+\frac{1}{a_2}\right) \left(\p{I}-\frac{\lambda}{4}\boldsymbol{\epsilon}\cdot \hat{\p{z}}\right)\nonumber\\ 
     &- \frac{1}{\rev{2}\pi\eta^\mathrm{e}L}
     \left(
     \rev{\p{I}} 
     -\frac{\lambda}{\rev{4}}\boldsymbol{\epsilon}\cdot \hat{\p{z}}  \right)\Bigg]\cdot \p{F}\label{eq:Ldyn},\\
    \dot{\p{R}} =& \frac{1}{12\pi \eta^\mathrm{e}}\left( \frac{1}{a_2}-\frac{1}{a_1}\right)\left(\p{I} - \frac{\lambda}{4}\boldsymbol{\epsilon}\cdot \hat{\p{z}} \right)\cdot \p{F}\label{eq:Rdyn}.
\end{align}
The first term in Eq.~\eqref{eq:Ldyn} results from $\eta^{\rm o}$-dependent Stokes' law~\cite{hosaka2023lorentz, ovconventions}, while the second represents the hydrodynamic interactions~\cite{yuan2023stokesian} and has been simplified by the torque-free condition.
Equation~\eqref{eq:Rdyn} shows that the presence of odd viscosity can affect the center of mass motion when $a_1\neq a_2$.
However, dimer's center of mass coordinate will be subject only to \rev{an oscillatory} motion without net displacement over one cycle of the \rev{reciprocal} deformation.
\rev{During such conformational changes, the dimers deform its shapes in a time-reversal manner, and therefore self-propelled motion does not occur, as in the case of classical fluids. 
This is known as Purcell's scallop theorem~\cite{purcell1977}.}
We therefore focus on Eq.~\eqref{eq:Ldyn}, in which the odd-mobility structure leads to an emergent rotational dynamics.
Our goal is to obtain the dynamical equations for the orientational dynamics of the vector $\p{n}(t)$ for an arbitrary prescribed actuation of the dimer size $L(t)$.

Taking the time derivative of the vector connecting the particles \p{L} formally gives
\begin{align}\label{eq:Lsph}
    \dot{\mathbf{L}} = \dot L \mathbf{n} + L \dot{\p n},
\end{align}
\rev{where the evolution of the unit vector is expressed in terms of the polar and the azimuthal angles, $\theta,\phi$, as $\dot{\mathbf{n}} = \dot \theta \hat{\boldsymbol{\theta}} + \dot \phi \sin\theta\hat{\boldsymbol{\phi}}$ with $\hat{\boldsymbol{\theta}} = \cos\theta \cos\phi \hat{\p{x}} + \cos\theta \sin\phi \hat{\p{y}} - \sin\theta \hat{\p{z}}$ and $\hat{\boldsymbol{\phi}} = -\sin\phi \hat{\p{x}} + \cos\phi \hat{\p{y}}$. 
In the coordinate system with the orthogonal bases $\{\mathbf{n},\hat{\boldsymbol{\theta}},\hat{\boldsymbol{\phi}}\}$, one finds $\dot{\p{L}}=(\dot L, L\dot \theta, L\dot \phi \sin\theta )$, and the torque-free condition is simply given by $\hat{\boldsymbol{\theta}} \cdot \p{F} = \hat{\boldsymbol{\phi}} \cdot \p{F} = 0 $.
We then find $\p{F}=(F_n,0,0)$ with $F_n = \p{n} \cdot \p{F}$.}
\rev{Expressing the velocity-force relation~\eqref{eq:Ldyn} in the spherical coordinates, we find that the polar angle evolution vanishes ($\dot \theta = 0$), while the dynamics of the length and the azimuthal angle are nonzero:
\begin{align}
    \dot L &= \left[\frac{1}{6\pi\eta^\mathrm{e}}\left(\frac{1}{a_1}+\frac{1}{a_2}\right) - \frac{1}{2\pi\eta^\mathrm{e} L}\right] F_n,
    \label{eq:Ldot}\\
    \dot \phi &= \frac{\lambda}{4L}\left[\frac{1}{6\pi\eta^\mathrm{e}}\left(\frac{1}{a_1}+\frac{1}{a_2}\right) - \frac{1}{2\pi\eta^\mathrm{e} L}\right] F_n.
   \label{eq:Lsin}
\end{align}
Comparing Eqs.~\eqref{eq:Ldot} and \eqref{eq:Lsin} simply gives}
\begin{align}
    \dot \phi &= \frac{\lambda}{4}\frac{\dot L}{L}
    \label{eq:phidot}
    ,
\end{align}
\rev{for $0<\theta<\pi$.
Note here that for the special initial orientations when the dimer aligns with the axis of odd viscosity $(\theta=0,\pi)$, the dimer does not exhibit any rotational motion $(\dot\phi=0)$.
This is because the anti-symmetric components driving the azimuthal rotation vanish when $\theta=0,\pi$ [see Eq.~\eqref{eq:Ldyn}].}
Moreover, the obtained dynamics is independent of the azimuthal angle $\phi$, which is required by the system's cylindrical symmetry around $\hat{\mathbf{z}}$~\cite{khain2022}.
\rev{Moreover, to this order, the dynamics of the azimuthal angle are proportional to the dynamics of the conformational speed of the dimer length.} 
It is obvious from the Eq.~\eqref{eq:phidot} that there is an emergence of the rotational dynamics due to the asymmetric mobility tensor, which couples the linear change in the interdomain distance $L(t)$ with motion transverse to it.
In a classical fluid with shear viscosity alone or equivalently when $\lambda \to 0$, $\dot\phi$ vanishes, as it should.
When the dimer orients in the plane perpendicular to the axis of odd viscosity ($\theta =\pi/2$), Eq.~\eqref{eq:phidot} still holds.

In the following, we provide an analytical solution of the dimer orientational dynamics without specifying the form of $L(t)$.
The angular evolution~\eqref{eq:phidot} can be rewritten as
\begin{align}
    \frac{d\phi}{dt} = \frac{\lambda}{4}\frac{d}{dt}\ln L ,
\end{align}
which can readily be solved
\begin{align}\label{eq:phisol}
    \phi(t) = \phi(0)+\frac{\lambda}{4}\ln\frac{L(t)}{L(0)} .
\end{align}
For a non-vanishing odd viscosity, the dimer undergoes oscillatory rotation about $\hat{\p{z}}$ as a function of $L(t)$.
When the dumbbell expands and the domains move apart, the dimer axis undergoes a transient anti-clockwise $(\lambda>0)$ \rev{[Fig.~\ref{fig:fig1}(b)]} or a clockwise precessional rotation $(\lambda<0)$, depending the sign of the odd viscosity.
The results in this section suggest that at small values of odd viscosity, the azimuthal angle exhibits the \rev{oscillatory} behavior as a function of $L(t)$, while the polar angle remains constant.

\subsection*{Higher order effects of odd viscosity}

We have shown that the odd viscosity, to leading order, gives rise to the precession dynamics of a dumbbell dimer about the odd viscosity direction $\hat{\mathbf{z}}$.
Here we extend the model to take into account higher order contributions and investigate whether the dynamics also exhibits the out-of-plane dynamics that includes the evolution of $\theta$.

\rev{By expanding $\dot{\p{L}}$ in Eq.~\eqref{eq:Ldotgen} up to the quadratic order in $\lambda$,  the distance between the spheres evolves as
\begin{align}
    \label{eq:nohydro}
    \dot{\p{L}} = 
    \frac{1}{6\pi\eta^{\rm e}}
    \left(
    \frac{1}{a_1}
    +
    \frac{1}{a_2}
    \right)
    \left[
    \left(
    1
    -\frac{\lambda^2}{10}
    \right)
    \mathbf{I}
    +
    \frac{\lambda^2}{20}
    \hat{\mathbf{z}}\hat{\mathbf{z}}
    -\frac{\lambda}{4}\boldsymbol{\epsilon}\cdot\hat{\mathbf{z}}
    \right]
    \cdot\mathbf{F}.
\end{align}}
Here we have assumed that each sphere experiences no hydrodynamic interaction from the other, i.e., $\p{M}_{\alpha\beta}=\p 0$, so as to focus on the higher-order effects of odd viscosity on the self-mobility of each sphere.
In the above equation, one can see that the even orders of odd viscosity affect only the diagonal terms of the hydrodynamic mobility, because the diagonal contributions must be independent of the ``handedness" of the fluid chirality.
Moreover, the presence of anisotropy due to the odd viscosity axis $\hat{\mathbf{z}}$ creates unequal forces between axial and transverse directions.

\rev{To proceed with the calculation, we employ an approach similar to the previous section, working in the frame of the spherical angular variables.}
As detailed in Appendix~\ref{sec:appA}, the angular dynamics is found as
\begin{align}
    \dot \theta &= -\frac{\lambda^2}{20}\frac{\dot L}{L}\cos\theta\sin\theta\label{eq:thetadotsecond} 
    ,\\
    \dot \phi &= \frac{\lambda}{4}\frac{\dot L} {L}
    \label{eq:phidotsecond},
\end{align}
\rev{where the latter is valid when $0<\theta<\pi$, as before.}
It is clear from Eq.~\eqref{eq:thetadotsecond} that the out-of-plane dynamics of the dimer is possible to the quadratic order or higher in $\lambda$, meaning that the odd viscosity anisotropy gives rise to the \rev{transient} alignment behavior toward the chirality axis, \rev{followed by the net oscillatory motion during one cycle of the deformation.}
Consequently, the resulting orientational dynamics shows the precessional motion as well as \rev{the oscillatory behavior with respect to the $\hat{\mathbf{z}}$-axis}, while it undergoes the monotonic expansion or contraction in its size $L(t)$.
Expressions~\eqref{eq:thetadotsecond} and \eqref{eq:phidotsecond} are the main results of this work, where we find that the presence of chirality inside the fluid induces three-dimensional rotational oscillations depending on the expansion-contraction of the dimer.
Similar alignment behavior may take place also in an anisotropic environment with a preferred direction~\cite{lintuvuori2017hydrodynamics}.
Nevertheless, the fluid chirality we consider originates purely from the fluid chirality and the obtained alignment mechanism is distinctive, as it does not rely on external mechanism, such as ambient torque coming from the minimization of elastic energy in the case of nematic liquid crystals.

Similar to the first-order analyses, we find the orientational dynamics of a dimer vanishes, $\dot \theta = \dot \phi =0$, when $\theta=0$ and $\theta=\pi$, while for $\theta = \pi/2$, the polar-angle dynamics vanishes $(\dot\theta=0)$, but the evolution of $\phi$ still obeys Eq.~\eqref{eq:phidotsecond}.
Additionally, the emergence of the quadratic $\lambda$-dependence in $\dot \theta$ is required from the symmetry of a force dipole as a polar vector.
The combination of the polar vector and the pseudovector associated with odd viscosity then suggests the invariance under $\hat{\p{z}} \to -\hat{\p {z}}$, from which one can deduce the even order of $\lambda$ for the polar angle evolution.
The symmetry argument here is consistent with the chirotactic response of a solitary self-propelled microswimmer/active particle in chiral environments~\cite{hosaka2024chirotactic}.

Since the polar angle is not determined by the azimuthal angle dynamics due to the symmetry of the system, Eq.~\eqref{eq:thetadotsecond} is readily integrated.
Rewriting Eq.~\eqref{eq:thetadotsecond} such that
\begin{align}
    \frac{d}{dt}\ln \tan\theta = -\frac{\lambda^2}{20}\frac{d}{dt}\ln L,
\end{align}
we have 
\begin{align}\label{eq:thetasol}
    \tan\theta(t) = \tan\theta(0)\left( \frac{L(0)}{L(t)}\right)^{\lambda^2/20},
\end{align}
while the solution to Eq.~\eqref{eq:phidotsecond} is given by Eq.~\eqref{eq:phisol}.
We can see from Eq.~\eqref{eq:thetasol} that the polar angle oscillates depending on the instantaneous ratio between $L(t)$ and $L(0)$.
When the dumbbell expands and the domains move apart, i.e., $L(0)/L(t)<1$, the dimer reorients itself along the $\hat{\p z}$-direction, \rev{meaning that the angle between $\p{n}$ and the axis of odd viscosity $\hat{\p{z}}$ becomes smaller}, while the contraction induces the \rev{the opposite effect, namely,} alignment in the plane perpendicular to $\hat{\p{z}}$ without any preferred direction \rev{[Fig.~\ref{fig:fig1}(c)]}.
This is because the systems has cylindrical symmetry about $ \hat{\p{z}}$ and the two-dimensional plane $(x,y)$ is isotropic, as can be also deduced from Eq.~\eqref{eq:phisol}.

\begin{figure}[t]
\centering
\includegraphics[width=1\linewidth]{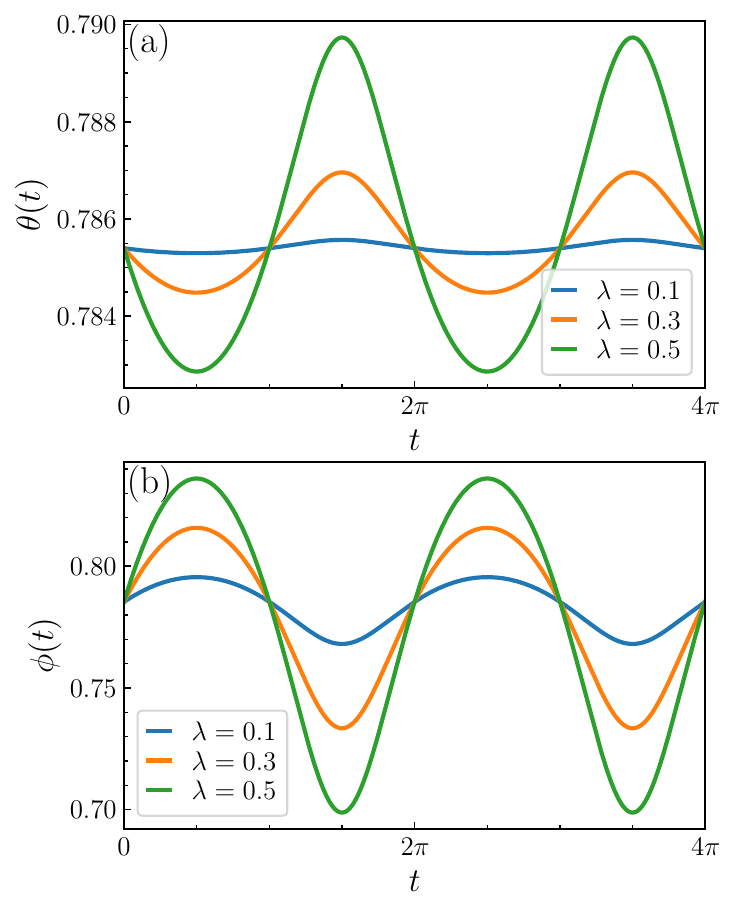}
\caption{\rev{
The angular dynamics of the dimer orientation as a function of time $t$ for various values of $\lambda$ when the self-mobility is taken into account up to the quadratic order in $\lambda$ and there is no hydrodynamic interaction between the domains.
The dimer is initially oriented with $\theta(0) = \phi(0) = \pi/4$ and its length is prescribed as $L(t)=L_0 + \ell_0 \sin(t)$ for $L_0 = 1$ and $\ell_0=0.5$.
The evolution of (a) the polar angle $\theta$ [Eq.~\eqref{eq:thetasol}] and (b) the azimuthal angle $\phi$ [Eq.~\eqref{eq:phisol}].
}
}
\label{fig:fig2}
\end{figure}

\rev{Figure~\ref{fig:fig2} shows 
the angular dynamics of a dimer for different values of $\lambda$.
The kinematic description of $L(t)$ is simply assumed to be $L(t) = L_0 + \ell_0 \sin(t)$, with $L_0$ being the average length of the dimer and $\ell_0$ being the maximum amplitude of oscillation.
As seen in Fig.~\ref{fig:fig2}(a), the polar angle shows an oscillatory behavior in time, where the expansion of the dimer reorienting it towards the $z$-axis, while the contraction makes it away from the direction to align in the $(x,y)$-plane.
The magnitude of the oscillation of the angle is enhanced with increasing $\lambda$, and this can be seen also for the time-evolution of $\phi$ [Fig.~\ref{fig:fig2}(b)].
}

\rev{
In the above analysis, we did not take into account the hydrodynamic interaction between the spherical domains.
Here we briefly examine the higher-order odd viscosity effect in the opposite regime, where the dynamics are governed solely by the hydrodynamic interaction~\eqref{eq:Mab}, without contributions from the mobility of a sphere~\eqref{eq:Maa}.
This regime can be justified when the spheres are sufficiently small to be treated as point particles.
Interestingly, from the full hydrodynamic interaction~\eqref{eq:Mab}, we observe that the dimer does not undergo any out-of-plane dynamics due to the absence of anisotropic terms involving $\hat{\mathbf{z}}\hat{\mathbf{z}}$.
This vanishing contribution to the $\theta$ dynamics holds for any value of $\lambda$, highlighting the crucial role of the particle self-mobility in the transient aligning behavior identified in Eq.~\eqref{eq:thetadotsecond}.
}

\rev{In Appendix~\ref{sec:appC}, we discuss another type of odd systems, the system with Lorentz forces, as a parallelism to the odd-viscous system we have studied.
We analyze the dynamics of a charged dimer in the presence of an external magnetic field.
We demonstrate that the observed three-dimensional dynamics occur in general in other odd systems by revealing the underlying connection between chiral fluids and magnetic systems.}

\section*{Orientational dynamics of passive dimer with thermal fluctuations}

The coupling between the linear conformational change prescribed by $L(t)$ and the orientational dynamics suggests that a passive dimer undergoing thermal fluctuations may also show the rotational diffusion, which will be affected by odd viscosity.
Whether it actually takes place requires an analysis of the conformational dynamics using a stochastic model.
In the previous literature, there exist various different stochastic dimer models taking into account passive and active noises.
Some of passive models for dimers include the Ornstein-Uhlenbeck (OU) process~\cite{uhlenbeck1930theory} or dimers with variable stiffness constant that is related to the recently proposed OU$^2$ model~\cite{Cocconi2024}. For active dimers, on the other hand, one can consider the active version of an OU (AOU) process in the presence of active thermal fluctuations~\cite{PhysRevE.100.022601,PhysRevE.103.032607} or dimers which are associated to molecular machines and undergo conformational oscillations through some non-equilibrium chemical process~\cite{Hosaka2020, Chatzittofi2024}.
To our knowledge, however, odd viscosity and the resulting asymmetric mobility tensor have not been studied in the overdamped Langevin system in the context of chiral active fluids (See Ref.~\cite{yasuda2022} for the discussion on underdamped systems).

In the following, we consider a passive dumbbell embedded in an odd-viscous fluid in the presence of fluctuations. 
The interdomain hydrodynamic interactions are again assumed to be negligible, to analyze the interplay between the anti-symmetric self-mobility and the ambient noise.
To describe the dynamics in the overdamped regime, we use the following Langevin equation
\begin{align}\label{eq:langevin}
    \dot{L}_a = \mathcal{M}_{ab} F_b + \xi_a,
\end{align}
with $a,b = \{\ell, \theta, \phi\}$. 
The first term represents the deterministic dynamics in the basis of spherical coordinates with the mobility tensor $\boldsymbol{\mathcal{M}}$, as defined in Eq.~\eqref{eq:Msph}, and the second is the stochastic noise that satisfies $\langle\xi_a(t)\rangle = 0$ and $\langle \xi_a(t) \xi_b(t') \rangle = k_{\rm B}T(\mathcal{M}_{ab} + \mathcal{M}_{ba}) \delta(t-t') \equiv 2D_{ab} \delta(t-t')$.
Here $k_\mathrm{B}$ is the Boltzmann constant, $T$ is the temperature of the system, and $D_{ab}$ is the symmetric positive-definite diffusion tensor that depends on $\lambda$.  
With the choice that only the symmetric components of the diffusion tensor enter in the noise variance, it follows that $D_{ab}$ obeys generalized Einstein's relation~\cite{PhysRevE.108.024602,johnsrud2024generalized}.

To make analytical progress we assume that $L(t) = L_0 + \ell(t)$ where $\ell(t) \ll L_0$ with  $L_0$ being the average length of the dimer and $\ell(t)$ obeying a stochastic process either continuously or discretely. 
This setup implies that the angular oscillation will be small in magnitude and thus similarly with the length one can assume $\theta=\theta_0 +\delta \theta(t)$ and $\phi = \phi_0 + \delta\phi(t)$. As a simple example, we consider the case where $\ell(t)$ follows an OU process. By utilizing Eqs.~\eqref{eq:langevin} and \eqref{eq:Msph}, the linearized Langevin dynamics are given by (see Appendix~\ref{sec:appB} for a derivation) 
\begin{align}
    \dot \ell &= -\mu_\ell k \ell + \xi_\ell\label{eq:elldot},
    \\
    L_0\delta\dot \theta &= -\mu_\theta k \ell + \xi_\theta,
    \label{eq:deltathetadot}
    \\
    (L_0\sin\theta_0)\delta \dot\phi &=-\mu_\phi k\ell + \xi_\phi
    \label{eq:deltaphidot},
\end{align}
where $-k\ell$ represents the force due to the conformational dynamics and the mobilities $\mu_\ell,\mu_\theta,$ and $\mu_\phi$ represent the mobilities of the corresponding velocities.
\rev{It is worth noting that the angular diffusivities, $D_{\theta \theta}$, and $D_{\phi\phi}$, are not related to $\mu_\theta$ and $\mu_\phi$. In particular, as shown in Appendices~\ref{sec:appA}-\ref{sec:appB}, $\mu_\theta = \mathcal{M}_{\theta \ell}$ and $\mu_\phi = \mathcal{M}_{\phi \ell}$, where for the noise strength we have, $D_{\theta \theta} = k_{\rm B} T \mathcal{M}_{\theta \theta}$ and $D_{\phi \phi} = k_{\rm B} T \mathcal{M}_{\phi \phi}$.}

\rev{We are interested in calculating the second moment, particularly the angular mean squared displacement given by $\langle \delta \theta^2(t) \rangle + \langle \delta \phi^2(t) \rangle$ at the early dynamics of $\mu_\ell kt \ll 1$, where our linearization scheme is valid}. 
By solving Eqs.~\eqref{eq:elldot}-\eqref{eq:deltaphidot} and calculating the corresponding correlators, \rev{the angular mean squared displacement} is found as (see Appendix~\ref{sec:appB} for a derivation)
\rev{
\begin{align}\label{eq:drot}
&\langle \delta \theta^2(t) \rangle + \langle \delta \phi^2(t) \rangle \nonumber\\
&\simeq \frac{2}{L_0^2}\left(D_{\theta\theta} + \frac{D_{\phi\phi}}{\sin^2\theta_0}\right)t
+k_\mathrm{B}T\frac{a_1+a_2}{6\pi\eta^\mathrm{e}a_1a_2}\frac{\lambda^2k^2 }{24L_0^2} t^3,
\end{align}
where the first term results from the rotational diffusion of the dimer and the second represents the odd viscosity-induced correction to it, which is cubic in time.}
The presence of the oddity in the system couples the orientational dynamics with the spatial conformations, leading also to the change in the rotational diffusion coefficient.
In other words, the rotational oscillations (resulting from the \rev{reciprocal} conformational changes of a dumbbell dimer) affect the diffusivity of the system.
Interestingly, this chirality-induced diffusion is similar to the case where reciprocal motion influences spatial diffusion~\cite{PhysRevLett.106.178101}.

\section*{Discussion and conclusion}

We have investigated the three-dimensional dynamics of a dumbbell dimer in a chiral active fluid with odd viscosity.
In stark contrast to the classical case, the presence of chirality can create orthogonal forces (as a result of off-diagonal couplings in the mobility tensor), which induces motion perpendicular to the dimer axis.
The oscillatory motion of a dimer gives rise to the oscillatory forces, which in turn leads to the periodic angular dynamics.
To the first order in $\lambda$, we showed that only the azimuthal orientation of the dimer is affected by odd viscosity, leading to the precession dynamics around the odd viscosity axis $\hat{\p{z}}$.
On the other hand, the dynamics of the polar angle becomes apparent in quadratic order. 
The obtained three-dimensional rotational dynamics provides an insight into the new odd phenomenology beyond the understanding of isotropy-compatible odd viscosity in two-dimensional experimental setups~\cite{soni2019odd}.
Furthermore, the observed coupling between the chirality and anisotropy would facilitate a number of new phenomenology in the analogous problems of odd diffusivity/mobility~\cite{LuigiMuzzeddu2025} in two-dimensional systems, which include the tracer dynamics~\cite{LuigiMuzzeddu2025}, in many-body systems~\cite{PhysRevLett.129.090601,PhysRevLett.132.057102}, and polymer models in magnetic fields \cite{Shinde2022}.

\rev{
In principle, modifications to the dimer dynamics are expected when higher-order contributions beyond the leading hydrodynamic interaction [Eq.~\eqref{eq:Mab}] are considered.
One possible extension is the Rotne–Prager–Yamakawa tensor, which includes the next higher-order term in the far-field approximation~\cite{rotne1969variational, yamakawa1970transport}.
However, extensions that incorporate both higher-order corrections and arbitrary odd viscosity values are rarely reported.
Although deriving a Rotne–Prager–Yamakawa-type hydrodynamic mobility for odd fluids is beyond the scope of this work, such corrections could provide valuable insights into dimer dynamics, particularly regarding the influence of dimer domains on the hydrodynamic interactions.
}

Although the obtained three-dimensional rotational dynamics are associated with precession and alignment behaviors relative to the chirality direction, these effects are transient. 
This is evident from the fact that a dimer returns to its initial orientation without net displacement after one cycle of deformation.
This observation, along with our previous finding on a solitary microswimmer~\cite{hosaka2024chirotactic}, suggests that odd viscosity induces the reorientation of linked sphere swimmers with a force dipole or time-averaged dipolar flow field.
In the case of the three linked spheres~\cite{najafi2004, golestanian2008}, swimmers with different radii generate a nonvanishing far-field dipolar flow and they are therefore expected to exhibit rotational motion due to odd viscosity.

In addition to the angular dynamics, its coupling to net translational motion would lead to interesting trajectories.
Such motile agents can be possible in a wide range of systems with multiple degrees of freedom, such as the above-mentioned three-sphere microswimmer~\cite{najafi2004, golestanian2008} and its extended  variations~\cite{yasuda2023generalized}.
For example, in the case of the three-sphere microswimmer, the active and coordinated motion of the arms leads to self-propulsion, as a result of the non-zero enclosed area in the phase-space coordinates. 
As a parallelism to this, in the presence oddity in the system, coordinated motion of the different angular variables could lead to self-rotational motion and possibly enhanced rotational diffusion. 
Such an idea could be tested through numerical simulations of three-dimensional polymer chains, i.e., chains of dimers as an extension of the classical Rouse model, and might be experimentally realized in systems of charged active polymers under the influence of an external applied magnetic field.

We point out that studying \rev{full non-linear} rotational dynamics can possibly lead to enhancement of the diffusion, depending on which model to be employed for describing \rev{the spatial oscillations}, and thus the resulting effect is model-specific.
For instance, in the recently reported OU$^2$ process, the length dynamics is affected by the stochastic dynamics of the stiffness constant~\cite{Cocconi2024}, whereas in the AOU process, the diffusivity changes due to the presence of active thermal fluctuations~\cite{PhysRevE.100.022601}. 
Similarly in the models for chemically induced conformational oscillations, the second moment of the displacements becomes a function of the chemical reaction coordinate~\cite{Chatzittofi2024}.

\begin{acknowledgments}
We thank E.\ Kalz and A.\ Vilfan for fruitful discussions.
We acknowledge support from the Max Planck Center Twente for Complex Fluid Dynamics, the Max Planck Society, the Max Planck School Matter to Life, and the MaxSynBio Consortium, which are jointly funded by the Federal Ministry of Education and Research (BMBF) of Germany.
Y.H.\ acknowledges support from JSPS Overseas Research Fellowships (Grant No.\ 202460086).
\end{acknowledgments}

\appendix
\section{Derivation of Eqs.~\eqref{eq:thetadotsecond} and \eqref{eq:phidotsecond}}
\label{sec:appA}

In the spherical coordinates, as already mentioned in the main text, the time-derivative of the interdomain vector is expressed as $\dot{\p{L}}=(\dot L, L\dot \theta, L\dot \phi \sin\theta )$ using the orthogonal bases $\{ \p{n},\hat{\boldsymbol{\theta}},\hat{\boldsymbol{\phi}}\}$ [see also Eq.~\eqref{eq:Lsph}].
The force-velocity relation then follows $\dot{\p{L}} = \boldsymbol{\mathcal{M}}\cdot \p{F}$, where the mobility tensor~\eqref{eq:Maa}, to the second order in $\lambda$, is given by
\begin{align}\label{eq:Msph}
    &\boldsymbol{\mathcal{M}} 
    = \frac{1}{6\pi\eta^\mathrm{e}}
    \left(\frac{1}{a_1}+\frac{1}{a_2} \right)
     \nonumber\\
&\times
\begin{pmatrix}
1-\frac{\lambda^2}{10} +\frac{\lambda^2}{20}\cos^2\theta  & -\frac{\lambda^2}{40}\sin2\theta& -\frac{\lambda}{4}\sin\theta \\
   -\frac{\lambda^2}{40}\sin2\theta & 1-\frac{\lambda^2}{10} +\frac{\lambda^2}{20}\sin^2\theta & -\frac{\lambda}{4}\cos\theta\\ \frac{\lambda}{4}\sin\theta & \frac{\lambda}{4}\cos\theta & 1-\frac{\lambda^2}{10}
\end{pmatrix}.
\end{align}
The mobility tensor $\boldsymbol{\mathcal{M}}$ can be decomposed into the symmetric and antisymmetric parts, where the off-diagonal components are responsible for the angular dynamics of the dumbbell dimer.
Using the equation of $\dot L$ to solve for $F_n$, and then substituting back in the equations of $\dot \theta$ and $\dot \phi$, we obtain the differential equations~\eqref{eq:thetadotsecond} and \eqref{eq:phidotsecond} accurate to $\mathcal{O}(\lambda^2)$. 
Notice that the anisotropy imposed by the fluid chirality is reflected in the structure of the mobility tensor in Eq.~\eqref{eq:Msph}.

\section{Connection with the Lorentz force}\label{sec:appC}

We compare the odd forces to the Lorentz force due to an external magnetic field.
The Lorentz force acting on a charged particle is given by $f^\mathrm{Lor}_i = \epsilon_{ijz} \dot r_j B$, where likewise with the axis of odd viscosity, we have set the magnetic field as $\p{B} = B \hat{\p{z}}$. 
By assuming that the two monomers have equal charges, it is straightforward to show that $\dot r_{j} (\delta_{ij} - \epsilon_{ijz}B) = f_i$. 
For an arbitrary value of $B$, the equations of motion are then expressed as
\begin{align}
    \dot \theta &= 
    -\frac{B^2\sin2\theta}{2+B^2(1+\cos2\theta)}\frac{\dot L}{L}\label{eq:thetaB},\\
    \dot \phi &= \frac{B}{1+B^2}\biggl(\dot \theta \tan\theta -\frac{\dot L}{L}\biggr),
\end{align}
which shares a similar structure with Eqs.~\eqref{eq:thetadotsecond} and \eqref{eq:phidotsecond}. 
The corresponding solutions are 
\begin{align}
    \tan\theta(t) \sin^\gamma2\theta(t)
    =\tan\theta(0)\sin^\gamma2\theta(0)\biggl( \frac{L(0)}{L(t)}\biggr)^{2\gamma},
\end{align}
for the polar angle with $\gamma=B^2/(2+B^2)$ and
\begin{align}
    \phi(t) = \phi(0)-\frac{B}{1+B^2}\ln \frac{L(t)\cos\theta(t)}{L(0)\cos\theta(0)},
\end{align}
for the azimuthal angle.
When $B\ll 1$, the behavior described by these solutions is along the same lines as Eqs.~\eqref{eq:thetasol} and \eqref{eq:phisol}, which are accurate to $\mathcal{O}(\lambda^2)$.

\section{Derivation of Eqs.~\eqref{eq:elldot}-\eqref{eq:deltaphidot}  and Eq.~\eqref{eq:drot}}
\label{sec:appB}

Applying the small oscillations in the spherical coordinates, such that $L(t) = L_0 +\ell(t)$, $\theta(t) = \theta_0 + \delta \theta(t),$ and $\phi(t) = \phi_0 + \delta\phi(t)$, as done in the main text, we find from the Langevin equation~\eqref{eq:langevin} the linearized velocity vector $(\dot\ell, L_0 \delta \dot\theta, L_0\sin\theta_0 \delta \dot \phi)$. 
For an OU process, the force just obeys Hooke's law such that $f_n = -k\ell$. 
Defining $\mu_\ell \equiv \mathcal{M}_{\ell\ell}$, $\mu_\theta \equiv \mathcal{M}_{\theta\ell},$ and $\mu_\phi \equiv \mathcal{M}_{\phi\ell}$ gives Eqs.~\eqref{eq:elldot}-\eqref{eq:deltaphidot} in the main text.

We next solve the equation for the interdomain distance $\ell$, as given by Eq.~\eqref{eq:elldot}.
The solution can be formally obtained as
\begin{align}
    \ell(t) =  e^{-\mu_\ell k t}\int^t_0 dt^\prime\, \xi_{\ell}(t^\prime) e^{\mu_\ell k t^\prime}.
\end{align}
\rev{where $\ell(0) = 0$. This implies that the correlator of the length is
\begin{align}
    \langle\ell(t') \ell(s') \rangle = \frac{D_{\ell \ell}}{\mu_\ell k}\left(e^{-\mu_\ell k |t'-s'|} -e^{-\mu_\ell k (t'+s')}\right).
\end{align}
We then solve Eqs.~\eqref{eq:deltathetadot} and \eqref{eq:deltaphidot} for $\delta \theta$ and $\delta \phi$, which gives their corresponding solutions
\begin{align}
\delta \theta(t) &= -\frac{\mu_\theta k}{L_0}\int^t_0 dt' \ell(t') + \frac{1}{L_0}\int^t_0 dt' \xi_\theta(t'), \\
\delta \phi(t) &= -\frac{\mu_\phi k}{L_0 \sin\theta_0} \int^t_0 dt' \ell(t') + \frac{1}{L_0 \sin\theta_0}\int^t_0 dt' \xi_\phi(t').
\end{align}}
With these, we calculate the correlators.
It is worth noting that to the second order in $\lambda$, the only correction comes from the dynamics of $\delta \phi$. 
More specifically, the second moments become
\begin{align}
	\langle \delta \theta^2(t) \rangle &= 2
    \frac{D_{\theta \theta}}{L_0^2} t +\mathcal{O}(\lambda^4)
    ,\\
	\langle\delta \phi^2(t)\rangle &= \frac{2D_{\phi\phi}}{(L_0\sin\theta_0)^2}t \nonumber\\&+ \frac{\mu_\phi^2 k^2}{(L_0\sin\theta_0)^2}\int^t_0 ds'\int^t_0 dt' \langle \ell(t') \ell(s') \rangle .
    \label{eq:b4}
\end{align}
\rev{By performing the integrals we find that
\begin{align}
\int^t_0 ds'\int^t_0 dt'& \langle \ell(t') \ell(s')\rangle =\nonumber\\ 
&\frac{D_{\ell\ell}}{(\mu_\ell k)^3}\left( 2 \mu_\ell k t + 4 e^{-\mu_\ell k t}-3 - e^{-2\mu_\ell k t}\right)\nonumber\\
&\simeq \frac{2D_{\ell \ell}}{3}t^3 + \mathcal{O}(t^4).
\end{align}
Substituting the result back in Eq.~\eqref{eq:b4}, at early times ($\mu_\ell kt\ll 1$), results in the leading correction} in Eq.~\eqref{eq:drot}. We note that $D_{\theta \theta}$ and $D_{\phi \phi}$ are smaller in the presence of odd viscosity, which resembles other odd systems, such as the suppressed rotational dynamics of a two-dimensional charged dimer in a magnetic field~\cite{Shinde2022}.

\bibliography{myref}

\end{document}